\documentclass[twocolumn,prl,showpacs]{revtex4-1}
\usepackage[colorlinks,bookmarks=false,citecolor=blue,linkcolor=red,urlcolor=blue]{hyperref}
\usepackage{graphicx,epstopdf,color}
\usepackage{amsfonts}
\usepackage{amsmath,amssymb,mathrsfs}
\usepackage{bm}
\def\bc{\begin{center}}
\def\ec{\end{center}}

\newcommand{\bs}[1]{\boldsymbol{#1}}

\newcommand{\vac}{\left|\,0\,\right\rangle}
\newcommand{\ket}[1]{\left|#1\right\rangle}

\renewcommand{\a}{\alpha}
\renewcommand{\b}{\beta}
\renewcommand{\c}{\gamma}
\newcommand {\ea}{\eta_{\alpha}}
\newcommand {\eb}{\eta_{\beta}}
\renewcommand {\ec}{\eta_{\gamma}}

\newcommand {\eab}{\bar\eta_{\alpha}}
\newcommand {\ebb}{\bar\eta_{\beta}}

\newcommand {\bSa}{\bs{S}_{\alpha}}
\newcommand {\bSb}{\bs{S}_{\beta}}
\newcommand {\bSc}{\bs{S}_{\gamma}}

\def\eg{\emph{e.g.}\ }

\allowdisplaybreaks[1]
\begin{document}
\title{A family of spin-$\bs{S}$ chain representations of SU(2)$_{k}$ Wess--Zumino--Witten models}
 \author{Ronny Thomale${}^1$}

 \author{Stephan Rachel${}^2$}

 \author{Peter Schmitteckert${}^{3,4}$}

 \author{Martin Greiter${}^{4,5}$}

 \affiliation{${}^1$Department of Physics, Stanford University, Stanford, CA 94305, USA}
 \affiliation{${}^2$Department of Physics, Yale University, New Haven, CT 06520, USA}
 \affiliation{${}^3$Institute f\"ur Nanotechnologie, KIT, 76344 Eggenstein-Leopoldshafen, Germany}
 \affiliation{${}^4$Theorie der Kondensierten Materie, KIT, Campus S\"ud, D 76128 Karlsruhe}
 \affiliation{${}^5$Institut f\"ur Festsk\"orperphysik, Postfach 3640, KIT, D 76021 Karlsruhe}

 \pagestyle{plain}

 \begin{abstract}
We investigate a family of spin-$S$ chain Hamiltonians recently introduced by one of us~\cite{Greiter11}. For $S=1/2$, it corresponds to the Haldane--Shastry model. For general spin $S$, we find indication that the low--energy theory of these spin chains is described by the SU(2)$_{k}$ Wess--Zumino--Witten model with coupling $k=2S$. In particular, we investigate the $S=1$ model whose ground state is given by a Pfaffian for even number of sites $N$.  We reconcile aspects of the spectrum of the Hamiltonian for arbitrary $N$
with trial states obtained by Schwinger projection of two Haldane--Shastry chains.
\end{abstract}



\maketitle

{\it Introduction}--- Conformal field theory (CFT) has significantly deepened our understanding of quantum spin chains and phase transitions in two-dimensional space-time~\cite{belavin-84npb333,cardy1987,DiFrancescoMathieuSenechal97}. In particular, it has provided a platform to distinguish different universality classes of critical spin chains according to \eg the power law parameters of multi-point spin correlation functions. Moreover, the central charge parameter $c$ of the CFT emerges in experimentally connected quantities such as specific heat~\cite{bloete-86prl742,affleck86prl746}. For critical SU(2) spin-$S$ chains, the conformal symmetry is supplemented by a Lie group symmetry and yields a low energy field theory description in terms of Wess-Zumino-Witten (WZW) models~\cite{wess-71plb95,witten84cmp455}. The WZW action consists of a non-linear $\sigma$ term plus the Wess-Zumino term, whose topological coupling factor $k$ is constrained to be integer and defines the level $k$ of the WZW$_k$ model~\cite{DiFrancescoMathieuSenechal97}. In the absence of broken continuous symmetries, WZW$_1$ is the generic low energy theory for half-odd-integer antiferromagnetic spin chains~\cite{affleck-87prb5291}, while integer spin chains show a Haldane gap~\cite{haldane83pl464}. 
Spin chains associated with WZW$_{k>1}$ models only appear at rare multi-critical points, and are of particular interest for integer spin chains where they can mark phase transitions between different gapped phases~\cite{takhtajan82pl479,babudjan82pl479,affleck-87prb5291}.

The $1/r^2$ Heisenberg spin-$1/2$ chain independently found by Haldane and Shastry~\cite{haldane88prl635,shastry88prl639} plays a special role among all other models related to WZW$_1$. While any finite size chain generally exhibits logarithmic corrections to the WZW long-wavelength limit depending on the system length $L=aN$ where $N$ is the number of sites and $a$ is the lattice constant, the Haldane--Shastry model (HSM) exactly obeys the WZW$_1$ scalings for any finite $L$.  (This is related to its enlarged Yangian quantum group symmetry~\cite{haldane-92prl2021}.)  Moreover, the exact HSM ground state wave function corresponds to the Laughlin 
wave function~\cite{laughlin83prl1395} in terms of bosonic spin flip particles at $\nu=1/2$ filling, and establishes the notion of fractional quantization and statistics of spinons as the fundamental excitations of quantum spin chains~\cite{haldane91prl937}. In this respect, it is the one-dimensional analogue of the chiral spin liquid~\cite{kalmeyer-87prl2095,schroeter-07prl097202}, where the concept of topological order was introduced in spin models~\cite{wen-89prb7387}.

It was realized recently that not only the Laughlin state, but a subset of the bosonic Read-Rezayi quantum Hall series~\cite{Read-99prb8084} can be generalized to a series of singlet spin $S$ wave functions at spin flip particle filling $\nu=S$~\cite{greiter-09prl207203}. 
For one spatial dimension, these states are constructed by a Schwinger boson projection technique~\cite{greiter02jltp1029} of $k$ copies of HSM ground states. The possibility of defining SU(2) invariant spin $S=k/2$ states related to the parafermionic CFT construction of the Read-Rezayi states is intuitive, as their current algebra reduces to the SU(2)$_k$ Kac-Moody algebra~\cite{Read-99prb8084} for the relevant fillings. As one particularly interesting member, the $k=2$ state corresponds to an $S=1$ Pfaffian spin state, which promises to establish the basis at which manifestations of non-Abelian spinons in spin chains can be investigated. 

From the combined view of state properties and low energy theory, it is then natural to ask whether these states may establish finite size representations of spin chains as the conformally invariant fixed point of WZW$_{k>1}$ in the same way as the HSM does for WZW$_1$. A step in this direction has been recently accomplished by one of us~\cite{Greiter11}, who introduced a family of Hamiltonians which singles out the spin $S$ chain states obtained by projection from $k=2S$ HS ground states as exact ground states. In this Letter, we further investigate these Hamiltonians. We numerically show that these $S=k/2$ spin chains are critical and indeed connected to the WZW$_k$ in the long wavelength limit. In contrast to the HSM, however, we find that the $S>1/2$ models exhibit logarithmic corrections and hence do not describe the conformally invariant fixed point of WZW$_{k>1}$. For the $S=1$ chain, we further analyze the excited states of the model, and generalize the Schwinger boson projection method to trial states for the simplest excitations.

\begin{figure}[t]
\begin{minipage}{0.99\linewidth}
\includegraphics[width=\linewidth]{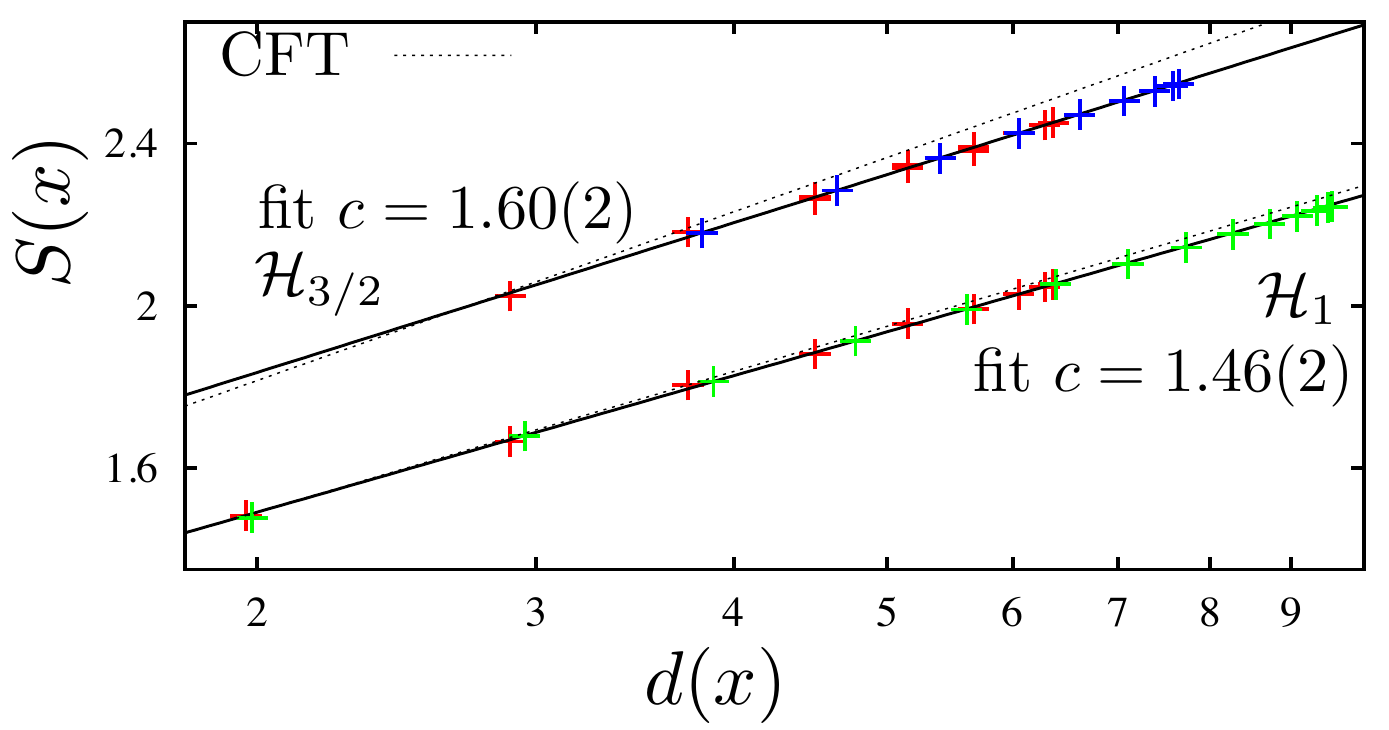}
\end{minipage}
\caption{(color online) Entanglement entropy $S(x)$ of $\mathcal{H}_S$ for $S=1$ and $S=3/2$ vs.\ the conformal length $d(x)=L/\pi \sin{\pi x/L}$. Data points are given for $L=20$ (red), $L=24$ (blue), and $L=30$ (green). The thermodynamic WZW results are sketched by dashed lines.  The fitted central charges (solid lines) agree well with SU(2) WZW$_{2S}$ field theory within reasonable error bars.}
\label{pic1}
\end{figure}

{\it Hamiltonians and ground states}--- The general Hamiltonian for the spin-$S$ chain~\cite{Greiter11} consists of a bilinear and biquadratic as well as a three-site Heisenberg term 

\begin{figure}[h]
\begin{minipage}{0.90\linewidth}
\includegraphics[width=\linewidth]{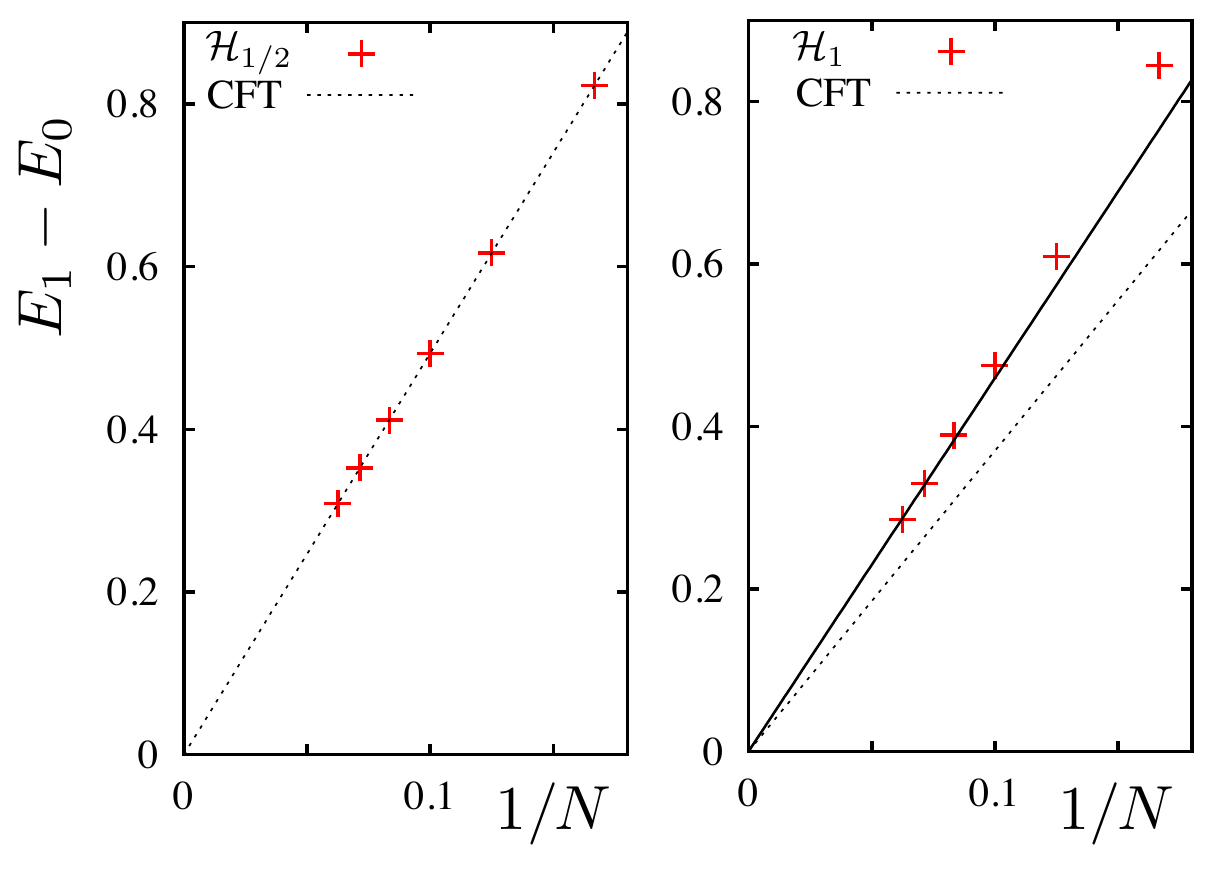}
\end{minipage}
\caption{(color online) Scaling dimensions for the Haldane Shastry model (left) and $\mathcal{H}_1$ (right). Left: The finite size data (crosses) exactly matches the estimated CFT result $x=\frac{1}{2}$ (dashed line). Right: The finite size linear fit (solid line) shows a mismatch with the estimated CFT result $x\sim \frac{3}{8}$.}
\label{pic2}
\end{figure}

\begin{widetext}
\begin{equation}
  \label{ham-s}
  \mathcal{H}_{S} =\frac{2\pi^2}{N^2}
  \Bigg[
  \sum_{\substack{\a\ne\b}} \frac{\bSa\bSb}{\vert\ea-\eb\vert^2}
  -\frac{1}{2(S+1)(2S+3)}\sum_{\substack{\a,\b,\c\\ \a\ne\b,\c}}
  \frac{(\bSa\bSb)(\bSa\bSc) + (\bSa\bSc)(\bSa\bSb)}{(\eab-\ebb)(\ea-\ec)}
  \Bigg]
\end{equation}
\end{widetext}
where $\bs{S}_{\alpha}$ is a spin-$S$ operator acting on site $\alpha$ and periodic boundary conditions are imposed by sites parametrized as complex roots of unity $\eta_{\alpha}=\exp{(i\frac{2\pi}{N}\alpha)}$, $\alpha \in 1,2,\dots,N$. (Note that
the biquadratic two-site term is contained in the three-site term in \eqref{ham-s}
as the special case $\beta=\gamma$.)
For $S=1$, this Hamiltonian has very recently been obtained independently through field theoretical methods by Nielson, Cirac, and Sierra~\cite{nielson-11arXiv:1109.5470}.
$\mathcal{H}_{1/2}$ is the HSM~\cite{haldane88prl635,shastry88prl639} (the three-site term trivially simplifies with the biquadratic term in this case). In Holstein-Primakoff representation, the ground state, which is the Gutzwiller wave function~\cite{gutzwiller63prl159}, takes the $\nu=1/2$ Laughlin form $\Psi_0^{\text{HS}}(z_1, \dots, z_M) =\prod_{i<j}^M (z_i - z_j)^2 \prod_{i=1}^M z_i$, where $M=\nu N$ and the $z_i$'s denote the coordinates of the spin flip operators acting on a spin polarized vacuum. The $1$st degree homogeneous polynomial multiplied with the squared Jastrow factor ensures that the state is both real and a spin singlet.

$\Psi_0^{\text{HS}}$ is the building block for all other spin-$S$ ground states. The spin-$S$ state is constructed by symmetrization of the spin flip coordinates of $k=2S$ identical copies of $\Psi_0^{\text{HS}}$. Technically, this can be accomplished conveniently in terms of auxiliary Schwinger boson creation operators $a^\dagger$ and $b^\dagger$, where the symmetrization effectively reduces to a multiplication of $k=2S$ copies $\ket{\psi_0^S}= \left(   \psi_{0}^{\text{HS}}[a^{\dagger},b^{\dagger}]\right)^{2S}\vac$~\cite{greiter-09prl207203}. The ground state energy yields $\mathcal{H}_S\ket{\psi_0^S}=E_0\ket{\psi_0^S}$ with $E_0 = -\frac{2\pi^2}{N^2}\frac{S(S+1)^2}{2S+3}\,\frac{N(N^2+5)}{12}$~\cite{Greiter11}.  The $S=1$ ground state of~\eqref{ham-s} is hence computed from the Schwinger projection of two HSM ground states and takes the explicit form $\Psi_0^{\text{Pfaff}}(z_1, \dots, z_M) =\text{Pf}(\frac{1}{z_i-z_j})\prod_{i<j}^M (z_i - z_j) \prod_{i=1}^M z_i$, where $M=N$ and $\text{Pf}(\frac{1}{z_i-z_j})=\mathcal{A}\left[\frac{1}{z_1-z_2} \frac{1}{z_3-z_4} \dots   \frac{1}{z_{N-1}-z_{N}}\right]$, and $\mathcal{A}$ is the antisymmetrization operator. Note that as the HSM singlet ground state requires $N$ even, the ground states for all $S\ge 1$ of~\eqref{ham-s} require $N$ even as well.  The case of $N$ odd will be analyzed below.

\begin{figure*}[t]
\begin{minipage}{0.90\linewidth}
\includegraphics[width=\linewidth]{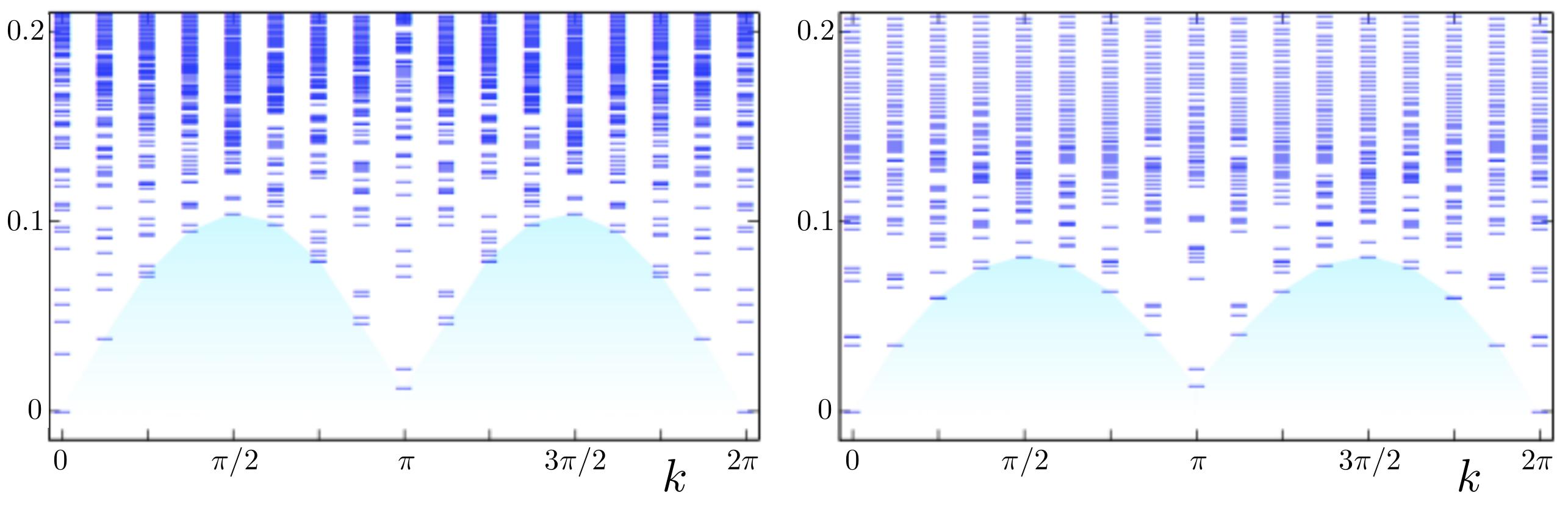}
\end{minipage}
\caption{(color online) $S=1$ energy spectra scaled to units of unity for the $S^z=0$ sector for $N=16$ sites. Left: $S=1$ Takhtajan-Babudjan Hamiltonian. Right: $\mathcal{H}_S$ for $S=1$.}
\label{pic3}
\end{figure*}

{\it Conformal field theory}--- Let us now determine the long wavelength universality classes of $\mathcal{H}_S$. We employ the density matrix renormalization group (DMRG)~\cite{white92prl2863} to compute the von Neumann entanglement entropy (EE) for large system sizes. To begin with, we checked the analytic formula for the ground state energy up to large values of $S$ and $N$. For gapped models, the EE saturates to a constant value; for gapless models, the EE diverges logarithmically, and, for finite chains takes the analytic form~\cite{calabrese-04jsmp06002}
\begin{equation}\label{ee}
S(x) \sim \frac{c}{3}\log{\left( \frac{L}{\pi}\sin{\frac{\pi x}{L}} \right)}.
\end{equation}
The computation of the ground states in~\eqref{ham-s} turns out to be challenging. One reason are the long-range terms in the Hamiltonian which adds to the enlarged local basis for higher spin $S$. For $S=1$ ($S=3/2$) we have kept up to 800 (1200) states of the reduced density matrix and performed up to $12$ ($14$) sweeps, which enabled us to keep the discarded entropy below $10^{-7}$ ($10^{-5}$). For \eqref{ham-s} with $S=1/2$, $1$, and $3/2$, we confirm critical behavior due to the log-divergent behavior of the EE. For $S=1/2$, $1$, and $3/2$, we find central charges $c=1$, $c=1.46(2)$ and $c=1.60(2)$ from chains up to $L=30$ [Fig.~\ref{pic1}], which are approximately consistent with the expected values $c_{\rm WZW}=3k/(k+2)$ for WZW$_k$.
The discrepancy between the numerically obtained $c$ and the asymptotically expected values is absent for $S=1/2$ but sets in for higher $S$. From finite size scaling, we cannot exclude the existence of non-monotonous corrections for small $L$ spin-$S$ chains. Still, over all we find numerical indication that the low--energy theory of $\mathcal{H}_S$ is the SU(2) WZW$_{2S}$ model within reasonable error bars.

{\it Logarithmic corrections}--- Having established that the CFT's related to the spin chains in~\eqref{ham-s} are of WZW$_k$ type, we now address the question of logarithmic corrections. As stated before, the HSM as the $S=1/2$ realization of \eqref{ham-s} shows no finite size corrections as compared to the long wavelength limit. To analyze this issue in more detail, we calculate the scaling dimension $x$ of the WZW$_k$ primary fields, as this quantity is highly sensitive to finite size corrections~\cite{fuehringer-08adp922}. Specifically, the value for the primary field at momentum $\pi$ can be extracted from $E_1-E_0=2\pi v x/L$, where $v$ is the velocity parameter and $E_0$ and $E_1$ are the energies of the ground state and first excited state.  Only small system sizes are needed to determine whether a spin chain resides at the conformally invariant fixed point or not. In Fig.~\ref{pic2} we have plotted $E_1-E_0$ for $\mathcal{H}_{1/2}$ and $\mathcal{H}_{1}$ as computed by exact diagonalization (ED) for chain lengths up to $N=16$. (As $H_1$ is not very sparse, the number of scattering elements is already of $\mathcal{O}(10^{13})$ for $N=16$.)  With $x=\frac{1}{2}$ and $\frac{3}{8}$ for $S=1/2$ and $1$ as predicted from CFT, we nicely observe the absence of finite size corrections for the HSM, but its presence for the $S=1$ chain [Fig.~\ref{pic2}]. This suggests that logarithmic corrections are present for $S>\frac{1}{2}$, and as a consequence that the models~\eqref{ham-s} do not represent the conformally invariant fixed points of WZW$_k$ for $k>1$.

{\it Spectrum of the $S=1$ chain}--- We investigate the structure of the finite size spectrum of \eqref{ham-s} for $S=1$ as obtained from ED. For this purpose, we compare it with the spectrum of the Takhtajan-Babudjan (TB) model~\cite{takhtajan82pl479,babudjan82pl479}, which is likewise connected to WZW$_2$ and describes the critical point between the dimerized phase and the Haldane gap phase of the bilinear-biquadratic spin-$1$ chain. The low energy spectra for $N=16$ are plotted in Fig.~\ref{pic3}. Aside from the singlet ground state at momentum $k=0$, the spin multiplet quantum numbers of the lowest energy modes in the different momentum subsectors are related. The lowest level at momentum $k=\pi$ determines the finite size gap. Both spectra look very similar, showing a two-lobe feature [Fig.~\ref{pic3}]. Still, the overlap of the Pfaffian and TB ground state is below $.85$ already for small system sizes, indicating that the models have different finite size structures. This becomes explicit as we develop a specific trial state Ansatz for the $S=1$ model in the following, which does not similarly apply to the TB model.   

\begin{figure*}[t]
\begin{minipage}{0.75\linewidth}
\includegraphics[width=\linewidth]{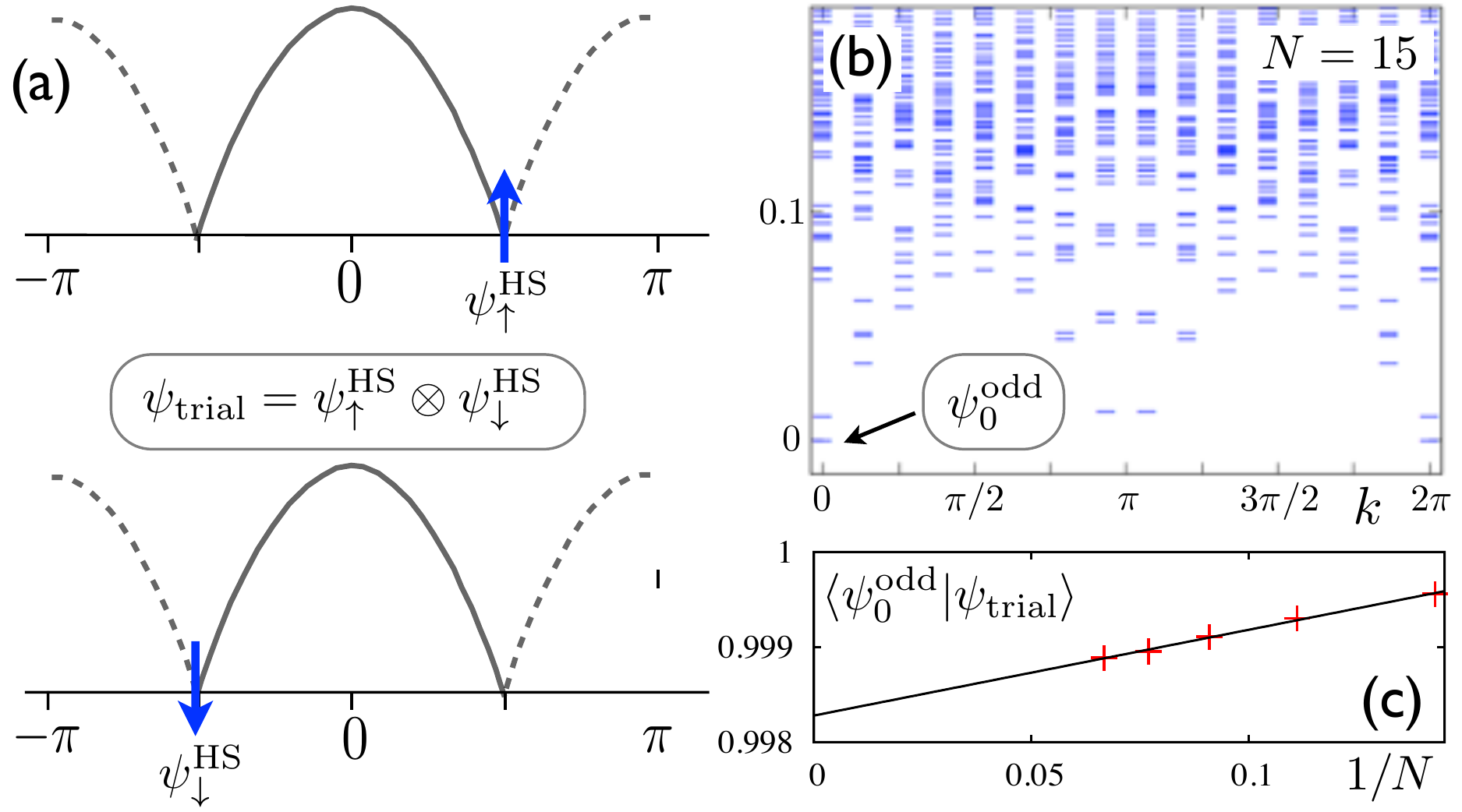}
\end{minipage}
\caption{(color online) (a) Construction of the $S=1$ trial state $\psi_{\rm trial}$ ($k=0$) for $N$ odd out of two Haldane-Shastry one-spinon states with momenta $k=\pm \pi/2$. (b) Spectrum of the $S=1$ chain for $N=15$. The ground state at $k=0$ is $\psi_0^{\rm odd}$. (c) Overlap $\langle \psi_{\rm trial} | \psi_0^{\rm odd}\rangle$ vs.\ $1/N$. For $N\to\infty$, the overlap is 0.9983(1).}
\label{pic4}
\end{figure*}

{\it Generalized Schwinger projection scheme}--- Up to now, we have only used our Schwinger projection scheme for HSM ground states to generate the exact $N$ even ground states for the spin-$S$ models in~\eqref{ham-s}. We now investigate what we can achieve when we employ the same approach for HSM excited states before projection. To begin with, we use the Schwinger projection scheme to obtain a suitable trial state for the $N$ odd ground state of the $S=1$ chain [Fig.~\ref{pic4}]. Consider first the two $S=1/2$ chains before projection. For $N$ odd, the low energy modes of the HSM are given by the single spinon branch which only covers one half of the momenta ($-\pi/2 < k_{\text{sp}} \leq \pi/2$ and $\pi/2 < k_{\text{sp}} \leq 3 \pi/2$ for $N=1\;\text{mod}\;4$ and $N=3\;\text{mod}\;4$, respectively) [Fig.~\ref{pic4}a]. 
The ground state for $N$ odd in the $S=1$ chain is located at $k=0$. ($N=15$ is shown in Fig.~\ref{pic4}b.) As phases of the wave function are preserved in Schwinger boson notation, they multiply under projection, and the total momentum of the projected wave function is given by the sum of single chain momenta before projection. In order to find a good trial state $\Psi_0^{\text{odd}}$ for the $S=1$ model \eqref{ham-s}, we choose the lowest energy one-spinon states in two HSM chains at $-\pi/2$ and $\pi/2$ before projection. To fully match the quantum numbers of the target state, we also project the spin part onto the singlet component of the projected two one-spinon wave functions. With this construction, we obtain an excellent overlap with the $N$ odd ground state of \eqref{ham-s} with $S=1$, which is of the order of $0.999$ and hardly changes with system size [Fig.~\ref{pic4}c].  Good overlaps can also be achieved for trial states to match other modes such as the lowest $N$ even eigenstates of the $S=1$ model in different momentum sectors [Fig.~\ref{pic3}].  There, we project the lowest lying two-spinon HS 
eigenstate for different momenta and a HS ground state together. This correspondence explains how the lobe features for the $S=1$ model are connected to the two-spinon levels of the HSM before projection.

From there, a unified picture emerges: we can interpret the Schwinger boson projection of Haldane-Shastry states as the creation of spinon product states. This is a peculiar property of the Schwinger boson projection of Haldane-Shastry chains, as we found elsewhere that taking two copies of a generic $S=1/2$ nearest neighbor Heisenberg model ground state as building blocks, for example, generates a trial ground state for a gapped spin-$1$ model, where the spinons become confined~\cite{rachel-09prb180420}.  In contrast to this, the HSM projection still provides gapless states, which is somewhat intuitive as the HSM is a free spinon gas. As the Hilbert space after projection is significantly smaller than the product space of the individual Haldane-Shastry chains before projection, this yields an overcomplete basis, and gives rise to selection rules that specify which many-spinons states before projection map onto each other after projection. This can be one way of defining the notion of a "blocking mechanism'' connected to the non-Abelian statistics of the spinons~\cite{Greiter11}. Comparing the trial states we constructed here via Schwinger boson projection with the actual eigenstates of \eqref{ham-s} for finite systems, we see that this construction 
yields reasonable approximations to the low energy modes of the system, even though it
does not provide us with exact eigenstates. The trial state we have constructed for the $S=1$ model with $N$ odd illustrates this point: since the spinons before projection have been chosen such that they cannot decay any further as they are located at the outer edges of the dispersion branches, we suppose that the "ideal" finite size WZW$_2$ spin chain model in the sense of spinon product states should exactly correspond to this construction and give an overlap of unity. The observed deviation from that is another way to interpret the logarithmic corrections we found for the Hamiltonian~\eqref{ham-s} for $S>1/2$.

\begin{acknowledgements}
RT thanks J.S.~Caux for comments and D.~P.~Arovas, B.~A.~Bernevig, and F.~D.~M.~Haldane for discussions as well as collaborations on related topics. MG thanks A.~W.~W.~Ludwig and K.~Schoutens for discussions. RT is supported by an SITP fellowship by Stanford University.
SR acknowledges support from DFG under Grant No. RA 1949/1-1. MG was supported by DFG-FOR 960.
\end{acknowledgements}


\end{document}